\documentclass{ifacconf}

\usepackage{graphicx}      
\usepackage{natbib}        

\usepackage{amsfonts}
\usepackage{amsmath}
\usepackage{amssymb}
\usepackage{array}
\usepackage{longtable}
\usepackage{comment}
\usepackage{subfigure}
\usepackage{xcolor}
\usepackage{upgreek}
\usepackage{color,soul}

\begin{document}
\begin{frontmatter}

\title{3D Map Reconstruction of an Orchard using an Angle-Aware Covering Control Strategy} 

\thanks[footnoteinfo]{This work was partially funded by the Italian IIT and MIUR within the 2017 PRIN (N. 2017S559BB) and by the Japan Society for the Promotion of Science (JSPS) KAKENHI under Grant N. 21K04104.}

\author[First]{Martina Mammarella}
\author[First]{Cesare Donati}
\author[Second]{Takumi Shimizu} 
\author[Second]{Masaya Suenaga} 
\author[Third]{Lorenzo Comba} 
\author[Third]{Alessandro Biglia}
\author[Fourth]{Kuniaki Uto} 
\author[Second]{Takeshi Hatanaka} 
\author[Third]{Paolo Gay} 
\author[First]{Fabrizio Dabbene}\footnote{corresponding author.}

\address[First]{Institute of Electronics, Computer and Telecommunication Engineering of the National Research Council of Italy, Corso Duca degli Abruzzi, 24, 10129, Turin, Italy (e-mail: {martina.mammarella@, cesare.donati@, fabrizio.dabbene@}ieiit.cnr.it).}
\address[Second]{School of Engineering, Tokyo Institute of Technology, Tokyo 152-8552, Japan (e-mail. {shimizu@hfg., suenaga@hfg., hatanaka@}sc.e.titech.ac.jp).}
\address[Third]{Department of Agricultural, Forest and Food Sciences – Università degli Studi di Torino, Largo Paolo Braccini, 2, 10095, Grugliasco, Italy (e-mail: {lorenzo.comba@, alessandro.biglia@, paolo.gay@}unito.it).}
\address[Fourth]{School of Computing, Tokyo Institute of Technology, Tokyo 152-8552, Japan (e-mail: uto@ks.c.titech.ac.jp).}

\begin{abstract}
In the last years, unmanned aerial vehicles are becoming a reality in the context of precision agriculture, mainly for monitoring, patrolling and remote sensing tasks, but also for 3D map reconstruction. In this paper, we present an innovative approach where a fleet of unmanned aerial vehicles is exploited to perform remote sensing tasks over an apple orchard for reconstructing a 3D map of the field, formulating the covering control problem to combine the position of a monitoring target and the viewing angle. Moreover, the objective function of the controller is defined by an importance index, which has been computed from a multi-spectral map of the field, obtained by a preliminary flight, using a semantic interpretation scheme based on a convolutional neural network. This objective function is then updated according to the history of the past coverage states, thus allowing the drones to take situation-adaptive actions. The effectiveness of the proposed covering control strategy has been validated through simulations on a Robot Operating System.
\end{abstract}


\end{frontmatter}

\section{Introduction}
\label{sec:intro}
In modern agriculture, the relevance of the role of Unmanned Aerial Vehicles (UAVs), also known as \textit{drones}, is rapidly growing (\cite{mammarella2021cooperationA}). Thanks to their enhanced capability to perform in-field operations in a precise and autonomous way, this typology of vehicles is leading to improvements in the context of the Agriculture 4.0 framework (\cite{radoglou2020compilation} \cite{mammarella2020cooperative}). As detailed in \cite{comba2019neural}, UAVs could allow to extend, both in terms of spatial and temporal dimensions, the capability to monitor the crop status during the whole growing season, thanks to light and transportable sensors. Within this context, UAVs are exploited, for examples, to reveal crop water stresses (\cite{guidoni2019method}), soil erosion (\cite{lima2021mapping}), fungal and pest infestations (\cite{calou2020use}), etc. Recently, the potential of 3D crop model informative content for agricultural applications has been investigated in \cite{comba20192d}, as an alternative to widely-exploited 2D maps (\cite{primicerio2015ndvi}). 3D map reconstruction through techniques like Structure-from-Motion (SfM) represents a powerful tool, even if it still represents a challenging task, due to the fact that the crop fields usually have poor and/or repetitive textures. Preliminary promising results have been achieved in the vineyard context where, thanks to a semantic interpretation approach, first proposed by \cite{comba2020semantic}, several relevant crop parameters, such as the leaf area index (see \cite{comba2020leaf}), have been remotely measured. To speed up and improve the data acquisition process, the use of fleets of drones is nowadays considered a promising solution. 

Classical covering control techniques typically requires drones to patrol over the environment by raising or lowering a density function (see e.g. \cite{sugimoto2015experimental}), according to the history of the past coverage states. On the other hand, to define and apply an \textit{effective} covering control strategy to the selected field, it is crucial to identify which areas are of higher relevance, e.g. crop canopy, and needs to be detected from  various viewing angles to improve the quality of the resulting map. This information can be resumed into a \textit{priority map}, describing the distribution of the importance index function over the field. Approaches similar to the one presented in \cite{comba2021semantic} can be exploited to automatically retrieve the 3D-points density function distribution according to the field characteristics, starting from multi-spectral maps.

The main drawback of available covering control strategies is the lack of situation-adaptive features, that does not properly adapt the control action to the relevance and observation rate of each region. This is due because the importance index is defined as a monotonically decreasing function. Hence, if a drone is approaching a well-observed region with low importance index, the objective function gets low and the standard gradient-based coverage schemes tend to decelerate the drones instead of making them quickly escape from the same region.

To overcome all the aforementioned limitations of the current covering control strategies and to properly customize and optimize the density function distribution to the orchard of interest, in this paper we combine the semantic interpretation approach presented in \cite{comba2021semantic} for deriving a priority map to the angle-aware covering control scheme described in \cite{shimizu2021} with the objective of providing an effective strategy to reconstruct a 3D map from rich viewing angles. The proposed scheme has been validated in simulation considering an apple orchard, first using a convolutional neural networks and an averaging filtering to identify the crop canopy and obtain the importance index field from a multi-spectral map, and then applying the distributed control strategy to a fleet of three UAVs. The preliminary results highlight the effectiveness of the covering controller and the adherence of the fleet behavior according to the evolution of the density function, demonstrating the capability of achieving situation-adaptive and angle-aware monitoring.

The remainder of the paper is structured as follows. In Section~\ref{sec:angle}, we present the selected covering control strategy to obtain a 3D map reconstruction using a fleet of drones. Then, Section~\ref{sec:framework} we describe the algorithm exploited to extrapolate from a given multi-spectral map the corresponding priority map. The preliminary results obtained exploiting the angle-aware covering control to an apple orchard are presented in Section~\ref{sec:results} while main conclusions are drawn in Section~\ref{sec:conlusions}.

\section{Angle-aware covering control}
\label{sec:angle}
\subsection{Preliminary definitions}
Let us consider a generic nonlinear dynamical system 
\begin{equation}
\label{sys}
\dot{x}=f(x)+g(x)u,
\end{equation} 
with $x\in\mathbb
R^n$, $u\in\mathbb{U}\subseteq\mathbb{R}^m$. The vector fields $f:\mathbb{R}^n\rightarrow\mathbb{R}$ and $g:\mathbb{R}^n\rightarrow\mathbb{R}^{n\times m}$ are assumed to be Lipschitz continuous, and the system \eqref{sys} has a unique solution $x(t)$ on $[t_0,t_1]$.
In this paper, we follow the approach in \cite{shimizu2021}, and we assume that the function $h:\mathbb{R}^n\rightarrow\mathbb{R}$ is a zeroing control barrier function (ZCBF) for the set $\mathbb{C}\doteq\{x\in\mathbb{R}^n\,|\, h(x)\geq 0\}$. It is shown in \cite{Ames2017} that any Lipschitz continuous controller $u(t)$ satisfying the constraint
\begin{equation}
    L_fh(x)+L_gh(x)u+\alpha(h(x))\geq 0,
    \label{eq:lie_constr}
\end{equation}
where $L_fh(x)$ and $L_gh(x)$ are the Lie derivatives of $h(x)$ along the vector field $f(x)$ and $g(x)$, and $\alpha$ is a class $\mathcal{K}$ function, renders the set $\mathbb{C}$ \textit{forward invariant}. 
Then, the control problem aims at determining the control action $u^*(x)$ closest to a given nominal input $u_{nom}$ while satisfying the forward invariance of $\mathbb{C}$, i.e.
\begin{equation}
\begin{aligned}
    u^*(x)\doteq \text{arg} &\min_{u\in \mathbb{U}}\,\,\, \|u-u_{nom}\|\\
    &\,\,s.t. \,\,\,\,\eqref{eq:lie_constr}. \nonumber
    \label{eq:QP_ctrl}
\end{aligned}
\end{equation}
\subsection{Problem settings}
\begin{figure}
    \centering
    \includegraphics[width=.7\columnwidth]{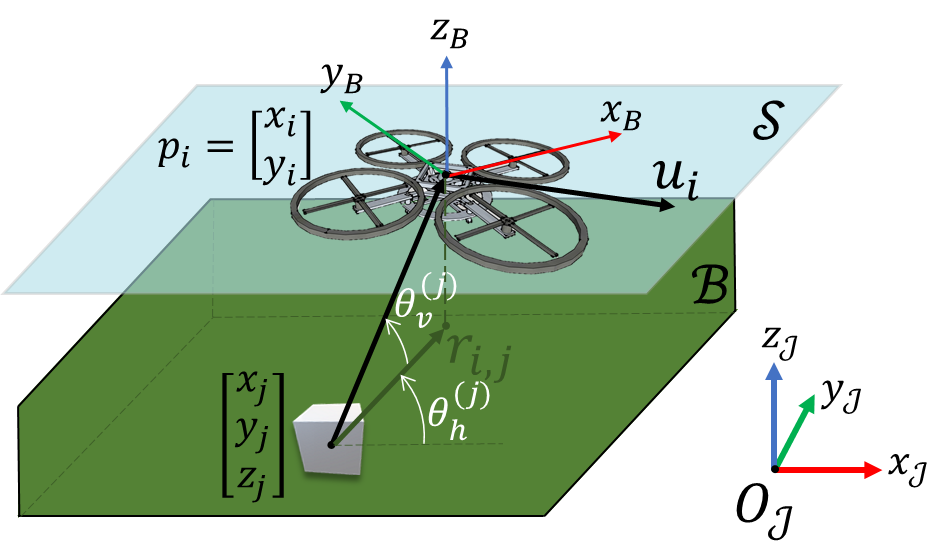}
    \caption{The $i$-th drone in $\mathcal{S}$ monitoring the $(\theta_h^{(k)},\theta_v^{(k)})$ side of the $j$-th object inside $\mathcal{B}$.}
    \label{fig:frame}
\end{figure}

The selected scenario involves $N$ UAVs, locally controlled such that they are characterized by common and constant altitude $z_c$ and attitude. This implies that the location of each $i$-th drone can be defined in a 2D space $\mathcal{S}$ as $p_i=[x_i,y_i]^\top$ with respect to an inertial frame $O_\mathcal{I}$ as represented in Fig. \ref{fig:frame}. Hence, the dynamics of each drone is defined as $\dot{p}_i= u_i,\,\forall i=[1,N]$, where $u_i\in\mathbb{U}\subseteq\mathbb{R}^2$ is the control velocity input to be designed.

We assume that the target scenario (i.e. the field to be reconstructed using SfM techniques) is contained into an a-priori known, compact set $\mathcal{B}\subset\mathbb{R}^3$, containing the ground surface. Then, the objective becomes to observe each point in the \textit{target field} $\mathcal{B}$ from rich viewing angles. This means that for the 3D map reconstruction we need to capture images of the target field $(x_j,y_j,z_j)\in\mathcal{B}$ from various $\theta_h^{(j)}$ and $\theta_v^{(j)}$ angles. In particular, $\theta_{h}^{(j)}\in[\pi,\pi)$ is defined as the horizontal angle and $\theta_{v}^{(j)}\in(0,\pi/ 2]$ is the vertical angle. Hence, the angle-aware coverage problem targets the coordinated region $\mathcal{Q}_c=\{q_{j}=[x_j,y_j,z_j,\theta_h^{(j)},\theta_v^{(j)}]^\top, \forall j\}$, which represents the target \textit{virtual field}. Then, given the mapping $\zeta:\mathcal{Q}_c\rightarrow\mathcal{S}$, we have 
\begin{equation}
q_{j} \mapsto \begin{bmatrix}x_j-(z_c-z_j)\tan{(\frac{\pi}{2}-\theta_v^{(j)})\cos{\theta_h^{(j)}}}\\
y_j-(z_c-z_j)\tan{(\frac{\pi}{2}-\theta_v^{(j)})\sin{\theta_h^{(j)}}}\end{bmatrix}.
\end{equation}
 The monitoring performance for the $i$-th UAV with respect to the point $q_{j}\in\mathcal{Q}_c$ is modeled by the distance between $p_i$ and the monitoring position $\zeta(q_{j})$. In particular, the performance function $\ell:\mathcal{P}\times \mathcal{Q}_c\rightarrow [0,1]$ is defined as
 \begin{equation}
     \ell(p_i,q_{j})\doteq \exp{\bigg(-\frac{\|p_i-\zeta(q_{j})\|^2}{2\sigma^2}\bigg)},
 \end{equation}
where $\sigma>0$ is a tuning parameter that depends on the sensor feature such that $\ell$ is small enough $\forall q_{j}\in\mathcal{Q}_c$.

\subsection{Objective function and controller design}
The next step consists in discretizing the 5D field $\mathcal{Q}_c$ into a collection of $M$ 5D cells, i.e. polyhedra of same area $A$, obtaining the new set $\mathcal{Q}\doteq\{q_\iota\}_{\iota\in[1,M]}$. Let us assign to each $\iota$-th cell an \textit{importance index} $\phi_\iota\in[0,\infty)$, which should decay if $q_\iota$ is monitored by one of the UAV and for which the decade rate depends on $\ell$. Then, we can define the following update rule for the importance index $\phi_\iota$ as
\begin{equation}
    \dot{\phi}_\iota(t)=-\delta\max_{i=[1,N]} \ell(p_i,q_\iota)\phi_\iota(t),\quad\phi_\iota(0)=\phi_\iota^{(0)}, 
\end{equation}
which renders each $\phi_\iota$ monotonically decreasing. Then, the control objective becomes to minimize an aggregate cost function $J\doteq \sum_{\iota=1}^M \phi_\iota  A$ to optimize the quality of the images collected by drones driving the cost $J$ towards zero. Moreover, to enhance the mission efficacy, a secondary objective is introduced to shape the UAVs behavior according to $\phi_\iota $:
\begin{itemize}
    \item the drone shall escape from region with small $\phi_\iota $, which corresponds to well-observed point $q_\iota$, the drone shall escape from this region;
    \item the UAV shall slow down and remain close to regions with large $\phi_\iota $.
\end{itemize}
This concept can be formalized as in \cite{shimizu2021} introducing a partition of the sampling set $\mathcal{M}$ as 
\begin{equation}
    \mathcal{V}_\xi(p)\doteq \{\iota\in\mathcal{M}\,|\, \|p_\xi-\zeta(q_\iota)\|\leq \|p_i-\zeta(q_\iota)\|, \forall i\in[1,N]\},
\end{equation}
and then describing the cost function rate as
\begin{equation}
    \dot{J}=\sum_{\iota=1}^m \dot{\phi}_\iota A=-\sum_{\xi=1}^N \mathcal{I}_\xi,
\end{equation}
with the metric $\mathcal{I}_\iota$ defined as $\mathcal{I}_\iota\doteq \sum_{\iota\in\mathcal{V}_\xi(p)} \delta \ell(p_\xi,q_\iota)\phi_\iota  A$.

This switching mode is enforced into the QP-based controller by taking: i) $h_\xi(p_i,q_{j})\doteq \mathcal{I}_\xi-\gamma$, with a given $\gamma>0$, as a candidate ZCBF; ii) $u_{nom}=0$; and iii) softening the constraints with the introduction of a slack variable $w_\xi$. The final QP problem becomes
\begin{equation}
    \begin{aligned}
    (u^*_\xi(x),w^*_\xi)= \arg &\min_{(u_\xi,w_\xi)\in \mathbb{U}\times \mathbb{R}} \,\, \varepsilon\|u_\xi\|^2+|w_\xi|^2\\
    &\,\, s.t.\,\,\, \dot{h}_\xi+\alpha(h_\xi)\geq w_\xi.\nonumber
    \end{aligned}
\end{equation}

This QP-based controller results hard to be implemented and solved in real time since the cardinality of $\mathcal{Q}_s$ tends to be very large. To address this issue, in \cite{shimizu2021}, the drone field $\mathcal{P}$ was discretized by a collection of $\ell$ polygons $\mathcal{A}_\ell$, all of the same area $A$, and corresponding gravity points $\mathcal{X}_\ell$. Then, according to the compression of $\mathcal{Q}$ onto $\mathcal{X}$ by the mapping function $\zeta$, in \cite{shimizu2021} the importance index $\phi$ was compressed onto $\psi_\ell \in [0, \infty)$ as
\begin{equation}
    \psi_\ell\doteq \sum_{\iota\in\mathcal{M}\,s.t.\, \zeta(q_{j})\in\mathcal{A}_\ell} \phi_\iota .
\end{equation}

Then, once the initial value of $\psi_\ell$ is defined, it can be updated without $\phi$ and the objective function $J$ can be further approximated as $J \approx \sum_{\ell} \psi_\ell A$. Consequently, the new QP-based controller can be derived similarly to \eqref{eq:QP_ctrl}. Further details can be found in \cite[Section V.A]{shimizu2021}.


\section{Selected framework}
\label{sec:framework}
\begin{figure}[!ht]
    \centering
\includegraphics[width=.9\columnwidth]{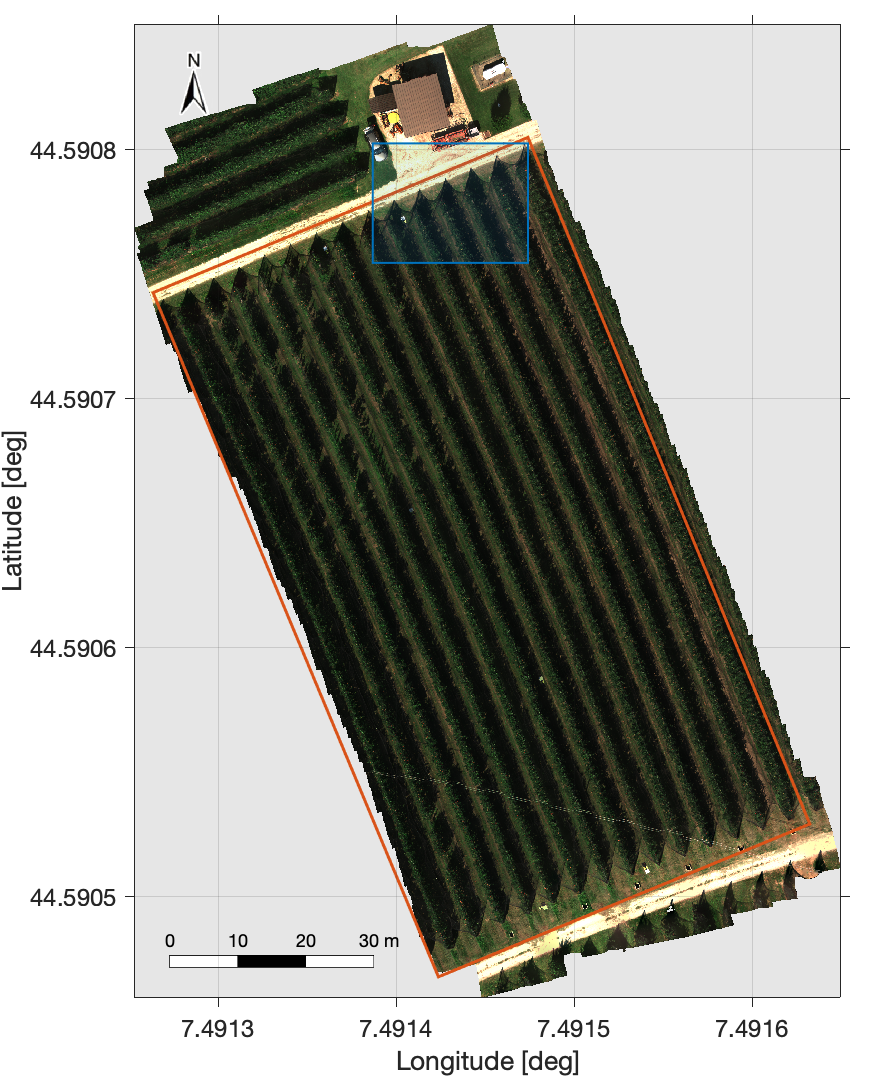}
    \caption{Aerial map of the selected apple orchard.}
    \label{fig:maps1}
\end{figure}

The case study selected envisions an apple orchard of about 0.4 ha located in Chiamina, Piedmont, North-west Italy (44°35’26” N, 7°29’30” E) (see Fig. \ref{fig:maps1}). Gala apples are cultivated in the selected orchard, and the distance between trees is about 0.9 m and the inter-row space, which is covered by dense grass, is 3.5 m wide. 

\begin{figure}[!ht]
    \centering
    \subfigure[ ]{\includegraphics[trim= .3cm .5cm 1.5cm .5cm, clip=true,width=.87\columnwidth]{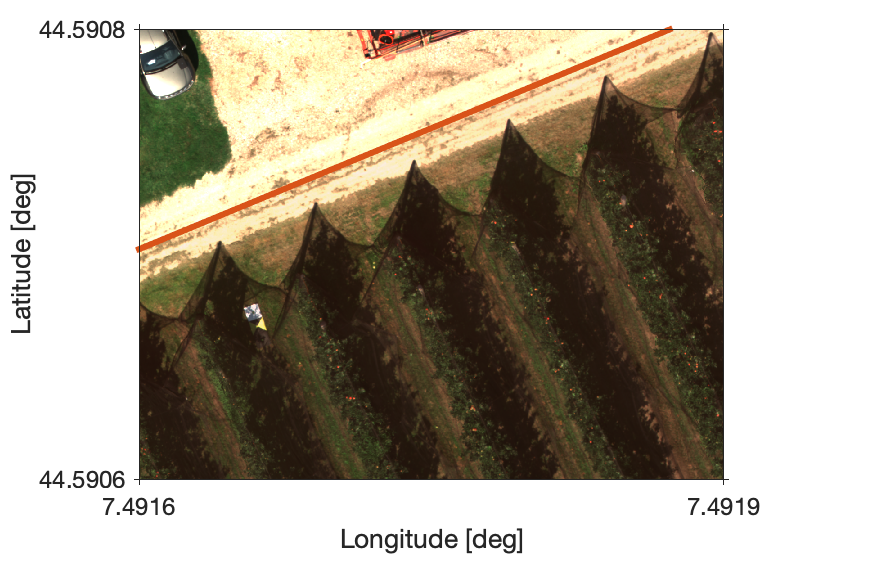}\label{fig:maps_a}}
    \subfigure[ ]{\includegraphics[trim= .3cm .5cm 1.5cm .5cm, clip=true,width=.87\columnwidth]{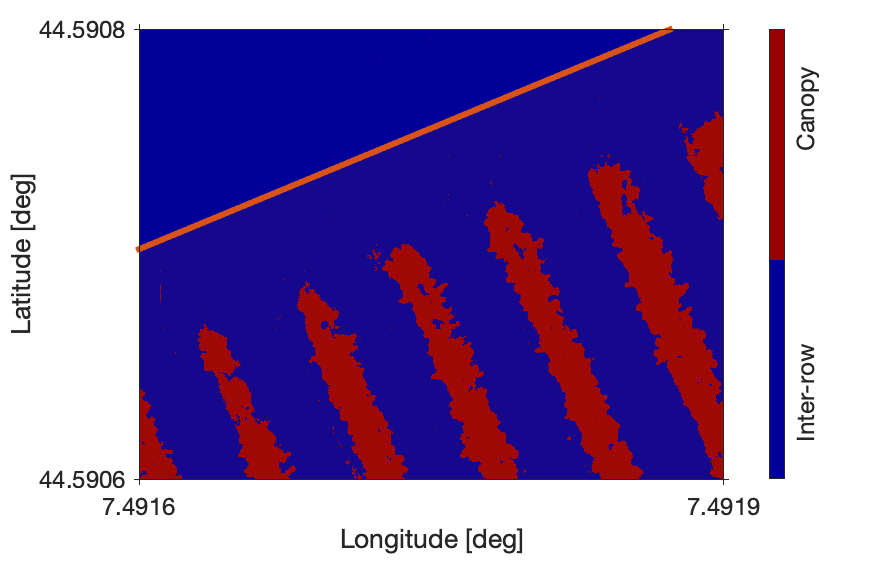}\label{fig:maps_b}}
    \subfigure[ ]{\includegraphics[trim= .3cm .5cm 1.5cm .5cm, clip=true,width=.87\columnwidth]{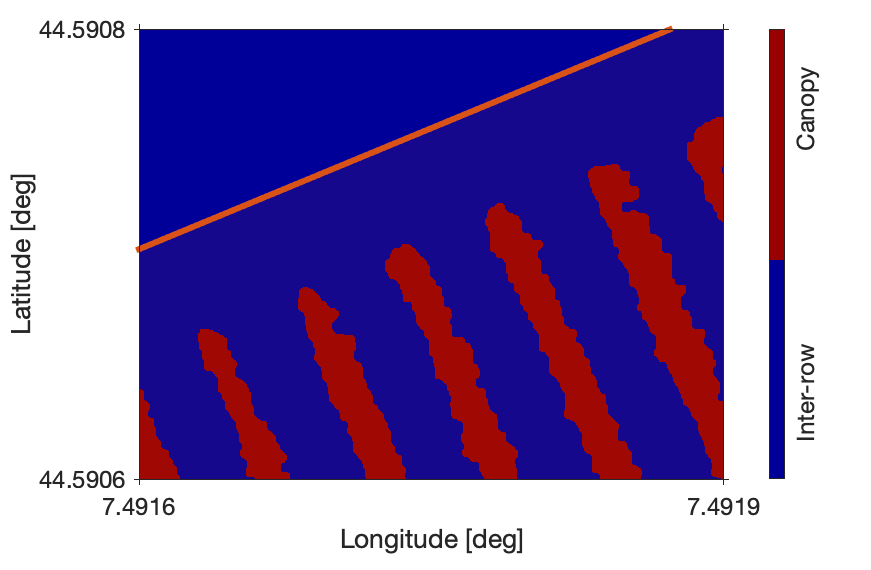}\label{fig:maps_c}}
    \subfigure[ ]{\includegraphics[trim= .3cm .5cm 1.5cm .5cm, clip=true,width=.87\columnwidth]{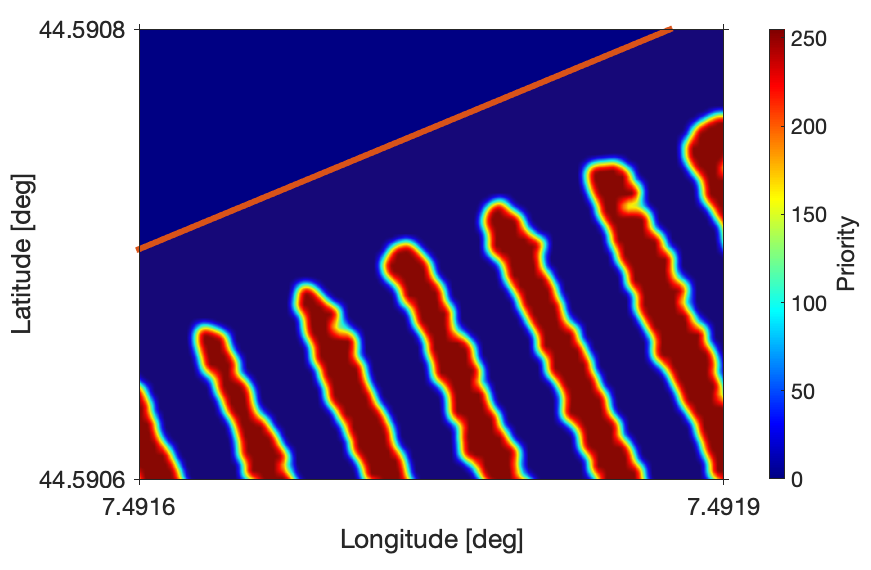}\label{fig:maps_d}}
    \caption{(a) Zoom-in of the orchard; (b) raw categorical map; (c) refined map; and (d) final priority map.}
    \label{fig:maps_zoom}
\end{figure}

To properly apply the proposed covering control strategy to the selected agricultural scenario, it is crucial to retrieve a specific importance index field $\phi^{(0)}$, adherent with the features of the orchard and the requirements related to the main objective of 3D map reconstruction. The distribution of $\phi^{(0)}$ over the field is named \textit{priority map}  and it is used to select regions of an orchard which are the target of the remote sensing task for 3D reconstruction. In particular, high priority regions will be those representing the crop canopy, while the rest of the map is considered less relevant.

In details, the priority map of the orchard was derived by processing a multi-spectral orthomosaics of the site, obtained by an aerial imagery campaign performed on September, 4th 2020. In particular, a MAIA S2 multi-spectral camera (SAL Engineering and EOPTIS) was mounted on a s900 DJI hexacopter with a GNSS receiver (u-blox 6, Pixhawk avionics). The UgCS Pro mission planning software (SPH Engineering) was used to define a set of waypoints, in order to maintain the height of the UAV flight close to 35 m with respect to the terrain and to guarantee a forward and side overlap between adjacent images greater than 80\%. The resulting ground sample distance (GSD) and the field of view (FOV), at 35 m of altitude, were equal to 1.75 cm pixel and 22.4x16.8 m, respectively. Image correction and true reflectance ratios calculation have been performed thanks to the incident light sensor (ILS), mounted on the top of the drone, which measure the ambient light level for each shot in each band. In addition, geometric correction, co-registration (or stitching) and radiometric correction was done by MAIA images software (MultiCam Stitcher Pro). Finally, the multi-spectral map of the whole orchard has been obtained by processing the image block with Agisoft Metashape (2020) software. Using the position of six ground markers (in-field determined with a differential GNSS), the map was also georeferenced in the WGS84 EPSG:4326 reference system.

\begin{figure}[!ht]
    \centering
\includegraphics[width=.9\columnwidth]{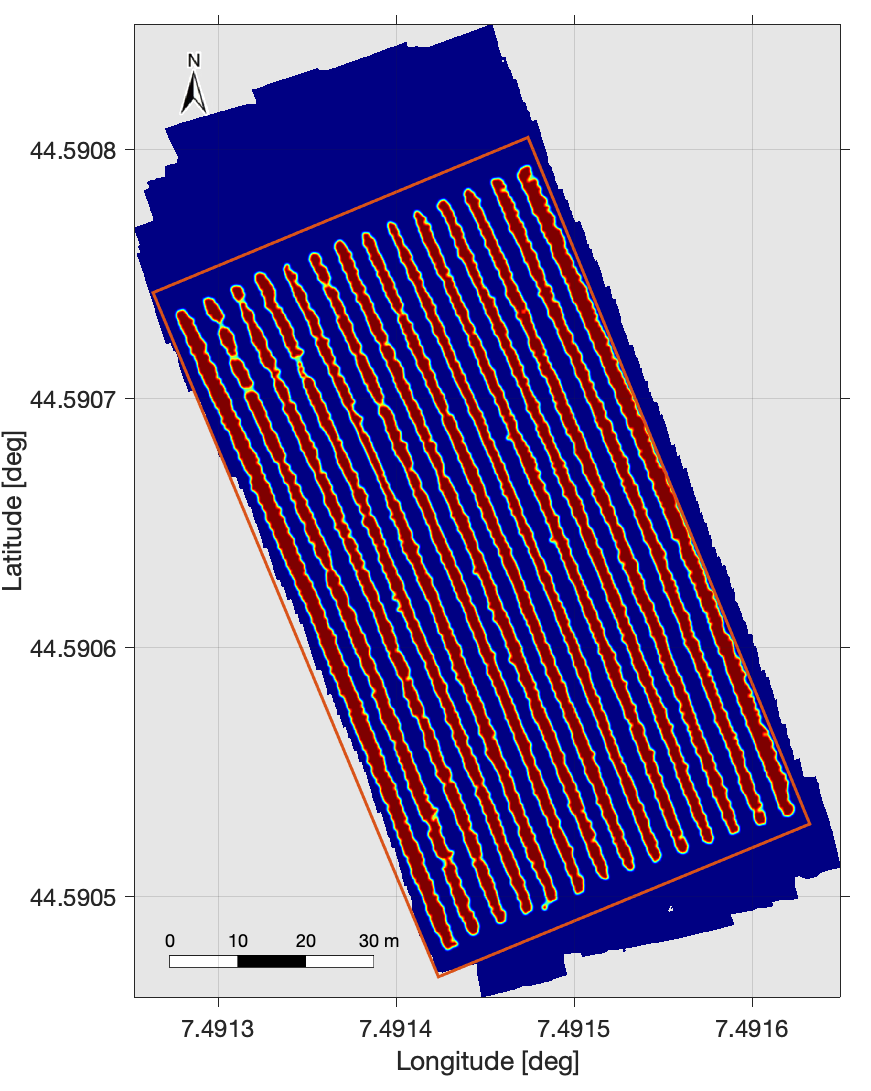}
    \caption{Priority map of the orchard, representing the distribution of the importance index $\phi^{(0)}$.}
    \label{fig:maps2}
\end{figure}

The procedure to retrieve the priority map from the multi-spectral ortho-mosaics  is based on 3 main steps: 1) the semantic interpretation by a convolutional neural network approach; 2) a refinement by morphological operations; and, finally, 3) a filtering and rescaling task. Pixels representing the crop canopy within the multi-spectral imagery are thus detected by a properly trained U-Net convolutional neural network (Mathworks Matlab®, 2020) and reported in a categorical map (see Fig. \ref{fig:maps_b}). To train, validate and test the U-Net, the ortho-mosaic was processed to select three different subsets of pixels, as described in \cite{comba2021semantic}. 
The raw categorical map provided by the U-Net (Fig. \ref{fig:maps_b}) was then processed with a sequence of morphological operators, in order to remove noise and small objects from the map, and to refine crop canopies boundaries (Fig. \ref{fig:maps_c}). In particular, the sub-sequentially performed operations are a closing and an opening operation, with a circular flat morphological structuring element with radius equal to 5 and 10 pixels, respectively. Then, to properly eliminate sharp gradient between clusters, an averaging filter with circular kernel was adopted (Fig. \ref{fig:maps_d}). The final priority map is represented in Fig. \ref{fig:maps2} and used as the initial distribution of the importance index $\phi^{(0)}$ for the covering control strategy as described in the following section.

\section{Numerical simulations}
\label{sec:results}
In this section, we describe the preliminary results obtained applying the proposed approach to an apple orchard. In the selected frameworks, we assume to cover the selected area with $N=3$ drones, whose initial positions were selected as $p_1=[5.0, 0.5]^\top$, $p_2=[7.5,0.5]^\top$, and $p_3=[10.0,0.5]^\top$, to cover the selected orchard. The local controller allowed to maintain a relative altitude of $10$ m with respect to the terrain while the UAV velocity is constrained into the input space $\mathbb{U}=\{u\in\mathbb{U}\,|\, \|u\|\leq 5\}$, limiting the drones acceleration to less than 5 m/s$^2$. The other controller parameters were set as summarized in Table \ref{tab:ctrl_set}. In particular, $\sigma=1$ allowed to drive $h\rightarrow 0$ at $3.5$ m,  which is the orchard inter-row space, while $\gamma=5000$ set the drone velocity around $2$ m/s when monitoring high-importance region.
\begin{table}[!ht]
    \centering
    \begin{tabular}{c|c||c|c||c|c}
    \hline
      Parameter & Value & Parameter & Value & Parameter & Value\\
      \hline
      $\sigma$ & $1$ & $\gamma$ & $5000$ & $\varepsilon$ & $10^{-4}$\\
      $d_{ca}$ & 0.5 & $\delta$ & 5 & $a$ & 5\\
      \hline
    \end{tabular}
    \caption{Covering controller settings.}
    \label{tab:ctrl_set}
\end{table}
 
\begin{figure*}[!ht]
    \centering
    \includegraphics[width=.69\paperwidth]{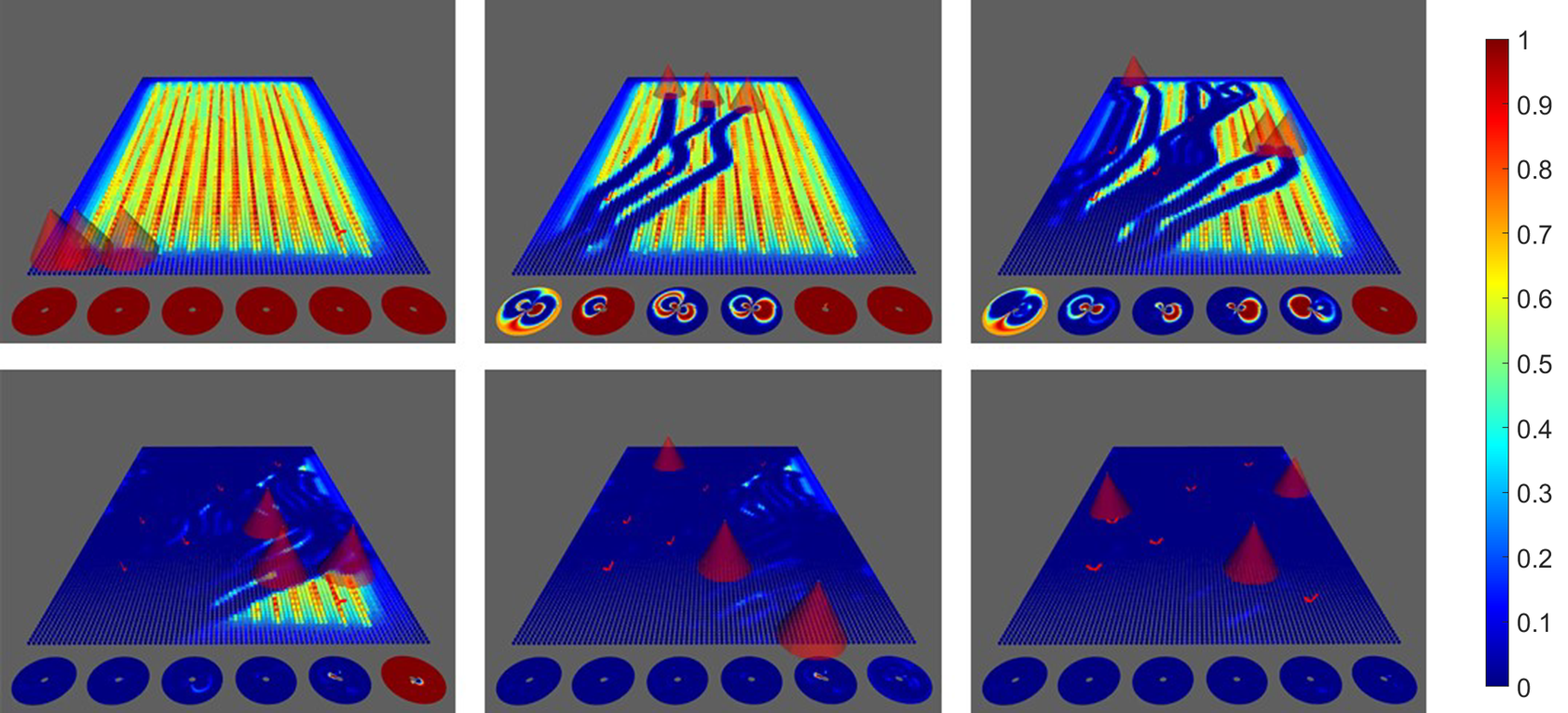}
    \caption{Evolution of the importance function $\psi$, where red circles denote the UAVs field of view. The circles at the bottom of each figure show the importance function $\phi$ with respect to viewing angles $(\theta_h,\theta_v)$ of the following coordinates triplets: $[10,20,0],  [10, 40,0],  [21, 30, 0],  [28, 60, 0],  [46, 80, 0],  [49, 10, 0]$.}
    \label{fig:snapshot}
\end{figure*}

From the priority map, retrieved as described in Section~\ref{sec:framework} and depicted in Fig. \ref{fig:maps2}, we can observe how the importance index $\phi^{(0)}$ is distributed over the orchard. Indeed, the semantic interpretation algorithm allowed to automatically identify the areas of higher interest, i.e. the tree rows, and to assign them a higher value of $\phi^{(0)}$ (yellow area) whereas we have much lower values of the importance index (blue areas) on the orchard boundaries and in the inter-rows zones. This information are later fed to the Python simulator, that requires a further down sample of the priority map into a $325\times200$ pixels representation (i.e. each pixel measures $0.3\times 0.3$ m) and then a compression of the map itself into a $80\times130$ points field. In details, we have that each cell of the virtual field $\mathcal{Q}$ is a $0.3$m$\times0.3$m$\times\frac{\pi}{30}$rad$\times\frac{\pi}{30}$rad polyhedron whose volume $A=\pi^2\times10^{-4}$ m$^2$rad$^2$, obtaining $m=1.95\times10^7$. About the drone field $\mathcal{P}$, it was discretized by $l=10,400$ $0.75\times0.77$ m$^2$ polygons $\mathcal{A}$.  


\begin{figure}[!ht]
\centering
\subfigure[ ]{\includegraphics[trim= 1.65cm 0cm 1.7cm 0cm, clip=true,width=.9\columnwidth]{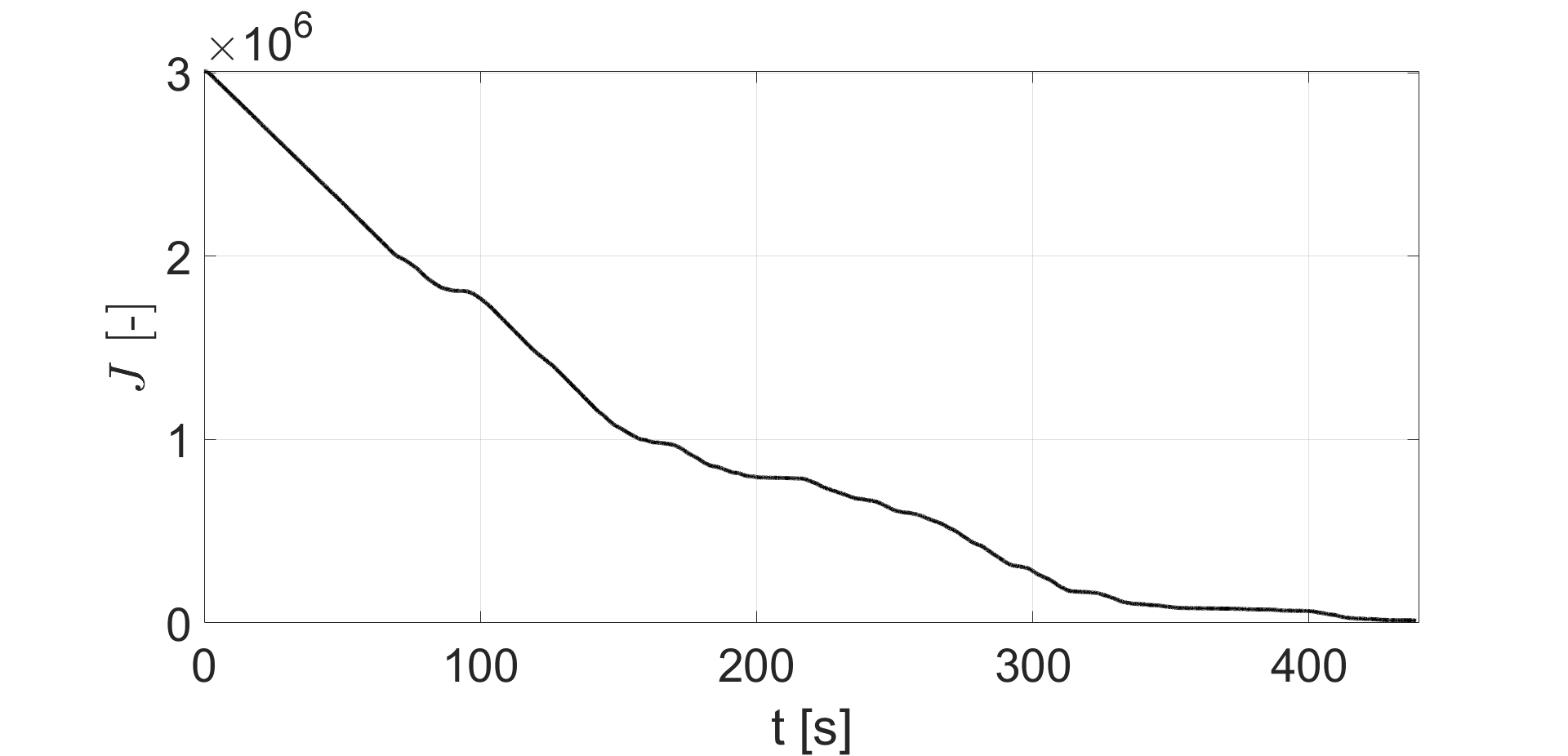}%
\label{fig:J}}
\hfil
\subfigure[ ]{\includegraphics[trim= 1.65cm 0cm 1.7cm 0cm, clip=true,width=.9\columnwidth]{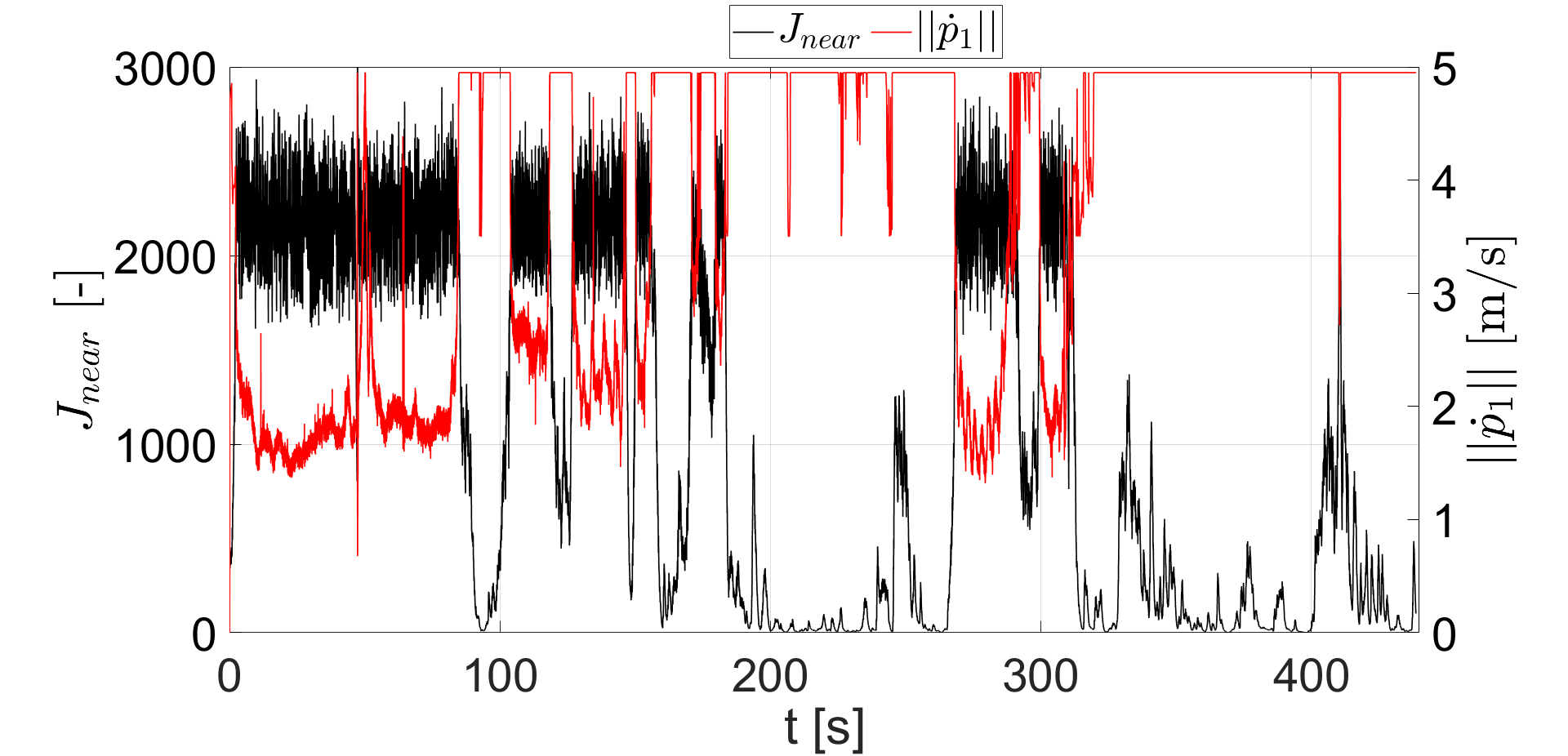}%
\label{fig:vJ}}
\caption{(a) Time series of the objective function $J$. (b) Time series of the UAVs’ velocity $\|\dot{p}_1\|$ (red) and $J_{near}$ (black).}
\label{fig:J_v}
\end{figure}

The next step consisted in validating the proposed covering control algorithm in a ROS environment, where CVXOPT was used to solve the QP of the proposed controller with an update frequency of $20$ Hz. In Fig. \ref{fig:snapshot}, we reported some frames from the simulation interface where it is possible to observe the evolution of the performance index $\psi$ according to the field coverage performed by the drones. In each frame, we can observe also the importance function $\phi$ with respect to the viewing angles for some specific check points. The drones took around $4$ minutes to cover the field and to collect the images from the selected viewing angles. In the second frame of the upper row of Fig. \ref{fig:snapshot} we can observe the collision avoidance strategy that makes the drones move away each other. On the other hand, in the third frame we can observe how the two drones on the right follows the priority map to collect images along the rows. Then, in the lower row of frames, it is possible to observe how progressively the drones covered the entire area from rich viewing angles, achieving the primary objective.

Fig. \ref{fig:J} shows the timeseries of the objective function $J$. It is possible to observe that $J$ is monotonically decreasing and converges to $0$. This implies that the UAVs would be able to collect \textit{ideal} images when $J\rightarrow 0$. Next, we needed to verify the secondary objective related to the adherence of the drone behavior according to the image sampling. In particular, we had to verify that the UAVs escape regions characterized by small $\phi$ (low importance or well-observed in the past) and to slow down when overflying high-relevance areas. From Fig. \ref{fig:vJ} we can observe that the UAVs average velocity when covering a high-relevance area was close to $2$ m/s whereas the airspeed was increased by the algorithm to 5 m/s when the parcel had low $\phi$. This can further easily verified introducing an additional performance parameter defined as the sum of importance index near the drone $p_1$, namely $J_{near}\doteq \sum_{\iota\in\mathcal{M}\,s.t.\,\|p_1-\zeta_\iota\|<2\sigma}\psi_\iota A$. The results depicted in Fig. \ref{fig:vJ} shows that the UAVs speed $\|\dot{p}_1\|$ tends to be small when the index $J_{near}$ is large (large $\phi$), and vice versa, thus highlighting the fulfillment of the secondary objective. 


\section{Conclusion}
\label{sec:conlusions}
In this paper, we applied a customized covering control strategy, that takes into account rich viewing angle to reconstruct 3D maps of an apple orchard. The proposed scheme is based on the preliminary extrapolation from a multi-spectral map of a priority field $\phi$ using a semantic interpretation approach. This allowed to determine the area of higher interest for the map reconstruction, coinciding whit the crop rows.
The effectiveness of the proposed scheme was finally demonstrated through numerical simulations.  
\bibliography{ifacconf}

\end{document}